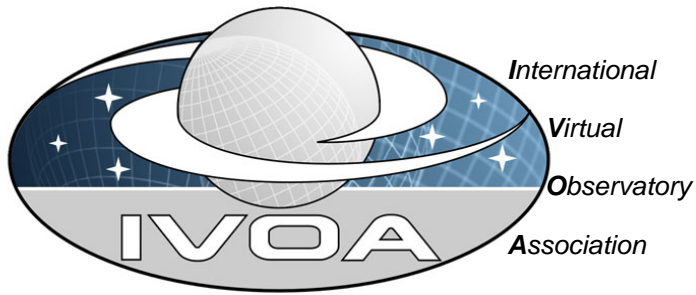

*I*nternational
*V*irtual
*O*bservatory
*A*ssociation

# TAPRegExt: a VOResource Schema Extension for Describing TAP Services
# Version 1.0

## IVOA Recommendation 27 August 2012

Working Groups:
   Registry WG, DAL WG
This version:
   http://www.ivoa.net/Documents/TAPRegExt-20120827
Latest version:
   http://www.ivoa.net/Documents/TAPRegExt/
Previous versions:
   http://www.ivoa.net/Documents/TAPRegExt/20110727
Authors:
   Markus Demleitner
   Patrick Dowler
   Ray Plante
   Guy Rixon
   Mark Taylor

## Abstract

This document describes an XML encoding standard for metadata about services implementing the table access protocol TAP [TAP], referred to as TAPRegExt. Instance documents are part of the service's registry record or can be obtained from the service itself. They deliver information to both humans and software on the languages, output formats, and upload methods supported by the service, as well as data models implemented by the exposed tables, optional language features, and certain limits enforced by the service.

## Status of this Document

*This document has been reviewed by IVOA Members and other interested parties, and has been endorsed by the IVOA Executive Committee as an IVOA Recommendation. It is a stable document and may be used as reference material or cited as a normative reference from another document. IVOA's role in making the Recommendation is to draw attention to the specification and to promote its widespread deployment. This enhances the functionality and interoperability inside the Astronomical Community.*

## Acknowledgments

This document has been developed with support from the German Astronomical Virtual Observatory (BMBF Bewilligungsnummer 05A08VHA).

This document has borrowed extensively from StandardsRegExt [SRE] working drafts.

## Syntax Notation Using XML Schema

This document defines the TAPRegExt schema using XML Schema [XSD]. The full schema document is listed in Appendix A. Parts of the schema appear within the main sections of this document; however, documentation nodes have been left out for the sake of brevity.

References to specific elements and types defined in the VOResource [VOR] schema include the namespace prefix, `vr`, as in `vr:Resource` (a type defined in the VOResource schema). References to specific elements and types defined in the TAPRegExt schema include the namespaces prefix, `tr`, as in `tr:TableAccess` (a type defined in the TAPRegExt schema). Use of the `tr` prefix in compliant instance documents is strongly recommended, particularly in applications that involve IVOA registries.

## Contents

## 1. Introduction

The Table Access Protocol TAP [TAP] allows VO clients to send queries to remote database servers and receive the results in standard formats. In addition, it defines means to discover database schemata on the remote side, to upload data from the local disk or third-party hosts, and more. TAP builds upon a variety of other standards, premier among which is the Universal Worker Service [UWS], which describes how client and server can negotiate the execution of a query and the retrieval of results without having to maintain a continuous connection.

To accommodate a wide variety of requirements, the TAP specification offers implementors many choices on optional features, resource limits, or locally defined functionality. One purpose of TAPRegExt is to allow the service to communicate such choices to remote clients using the mechanisms laid down in the VO Service Interfaces standard [VOSI].

Clients also need to discover TAP services offering certain kinds of data. Central to this is the concept of a registry in which resources can be described and consequently discovered by users and applications in the VO. Registries receive resource descriptions as defined in the IVOA standard [VOR]. In this schema, support for a standard service protocol is described as a service's capability; the associated metadata is contained within the service resource description's `<capability>` element.

TAPRegExt defines this capability element for TAP services. In the context of registering TAP services, an important role filled by TAPRegExt is the communication of supported data models to the registry.

## 1.1. TAPRegExt within the VO Architecture

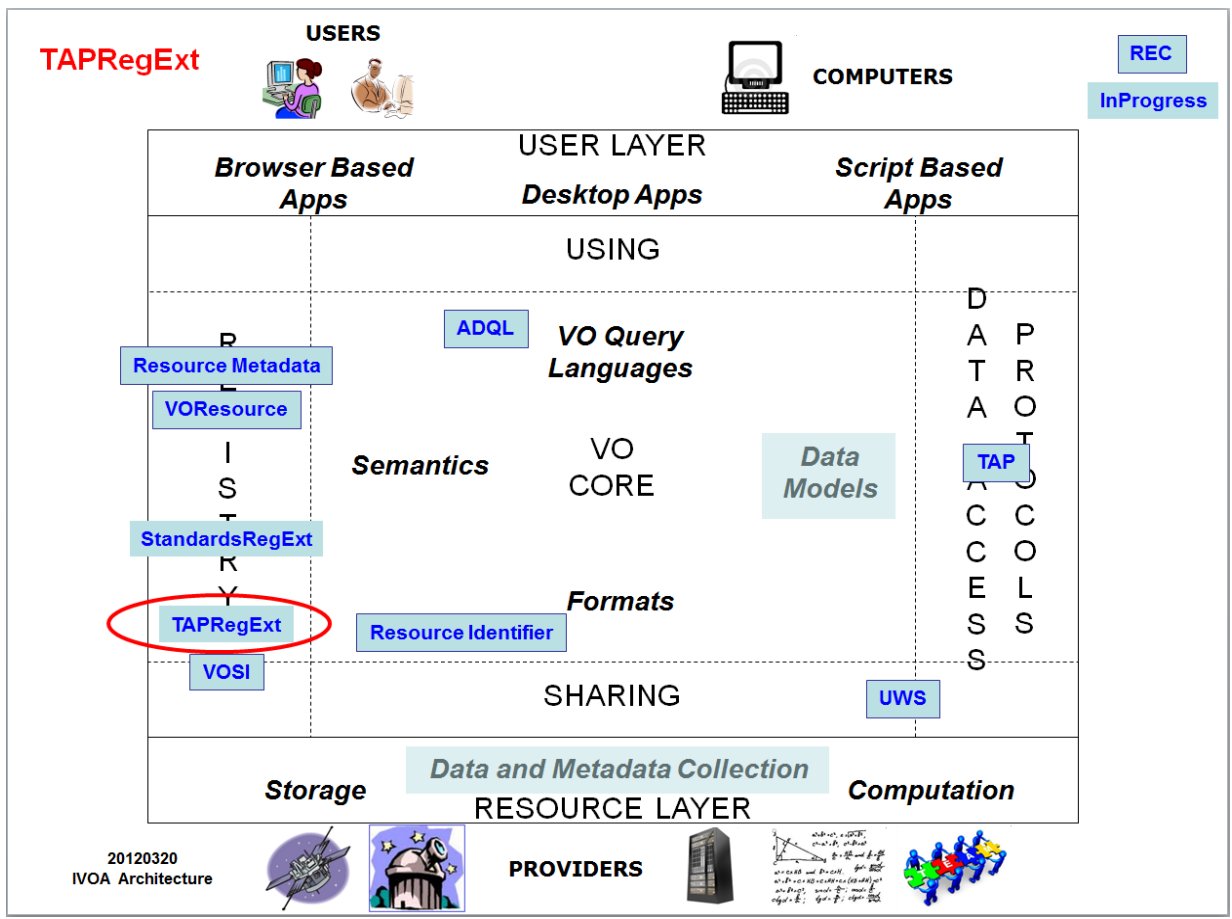

Figure 1: IVOA Architecture diagram with TAPRegExt and the related standards marked up.

This specification directly relates to other IVOA standards in the following ways:

VOResource, v1.03 [VOR]
    Descriptions of services that support TAP are encoded using the VOResource XML schema. TAPRegExt is an extension of the VOResource core schema.
TAP, v1.0 [TAP]
    The TAP standard defines some of the concepts that TAPRegExt deals with. The TAP standard document indirectly refers to this document in the specification of its capabilities endpoint.

UWS, v1.0 [UWS]
> The UWS standard describes additional parameters the choices of which are communicated using TAPRegExt.

StandardsRegExt [SRE]
> TAPRegExt uses the StandardKeyEnumeration mechanism introduced in StandardsRegExt to define controlled vocabularies.

This standard also relates to other IVOA standards:

IVOA Support Interfaces, v1.0 [VOSI]
> VOSI describes the standard interfaces to discover metadata about services; this document defines the response TAP services should provide on the `capabilities` endpoint described by VOSI.

IVOA defined data models
> Data models specified by the IVOA can define the structure of database tables holding instances of those data models. The first example of such a definition is given by [ObsCore]. Services providing access to such tables can declare that fact within TAPRegExt instance documents.

# 2. The Extension

## 2.1. The Schema Namespace and Location

The namespace associated with TAPRegExt VOResource extensions is `http://www.ivoa.net/xml/TAPRegExt/v1.0`. Just like the namespace URI for the VOResource schema, the TAPRegExt namespace URI can be interpreted as a URL. Resolving it returns the XML schema document (given in Appendix A) that defines the TAPRegExt schema.

Authors of VOResource instance documents may choose to provide a location for the VOResource XML schema document and its extensions using the `xsi:schemaLocation` attribute. While generators are free to provide any schema location (e.g., a local mirror), this specification recommends using the TAPRegExt namespace URI as its location URL (as illustrated in the example above), as in,

```
xsi:schemaLocation="http://www.ivoa.net/xml/TAPRegExt/v1.0
                    http://www.ivoa.net/xml/TAPRegExt/v1.0"
```

Note that you must give the `xsi:schemaLocation` of the TAPRegExt schema when the capability defined here is part of a published registry resource record as per the IVOA Registry Interface standard [RI]. This does not apply to the use in a TAP server's capabilities endpoint.

## 2.2. Declaring Instantiated Data Models

The IVOA defines certain data models that can be instantiated in database tables exposed by a TAP service. This allows a query built exclusively on a data model or a set of data models to work on all TAP services exposing tables instantiating the data model(s).

In TAPRegExt, a data model is identified by its IVORN [IVORN]. The first example for such a data model is ObsCore [ObsCore].

tr:DataModelType Type Schema Definition

```
<xs:complexType name="DataModelType" >
  <xs:simpleContent >
    <xs:extension base="xs:token" >
      <xs:attribute name="ivo-id" type="vr:IdentifierURI" use="required" />
    </xs:extension>
  </xs:simpleContent>
```

```
</xs:complexType>
```

| tr:DataModelType Attributes | | |
|---|---|---|
| Attribute | Definition | |
| ivo-id | *Value type:* | an IVOA Identifier URI: vr:IdentifierURI |
| | *Semantic Meaning:* | The IVORN of the data model. |
| | *Occurrences:* | required |

## 2.3. Languages Supported

TAP services may offer a variety of query languages. In TAPRegExt, the `language` element allows the communication of what languages are available on a service. TAP defines values of the `LANG` parameter to have either the form `<name>-<version>` or the form `<name>`, where the latter form leaves the choice of the version to the server. Therefore, a language is defined using a name and one or more versions.

The recommended way to associate larger amounts of documentation with a language entry in a capability element is via registration of the language using the mechanisms defined in [SRE] and associating the registry record with the language element through the latter's ivo-id attribute. The IVORN for the only language mandatory for TAP services, ADQL 2.0, is `ivo://ivoa.net/std/ADQL#v2.0`.

The type of the `ivo-id` attribute on version is `xs:anyURI` as opposed to `vr:IdentifierURI` since the latter does not allow frament identifiers. This is fine when referring to complete standards (as for the `dataModel` above), but will not work when the entities referred to are members of, say, StandardsRegExt enumerations. Hence, we allow any URI in the schema for this attribute. The description constrains the value to be an IVORN, though. The same reasoning applies to the `ivo-id` attributes of `outputFormat` and `uploadMethod`.

tr:Language Type Schema Definition

```
<xs:complexType name="Language" >
  <xs:sequence >
    <xs:element name="name" type="xs:NCName" />
    <xs:element name="version" type="tr:Version" minOccurs="1"
               maxOccurs="unbounded" />
    <xs:element name="description" type="xs:token" minOccurs="0" />
    <xs:element name="languageFeatures"
               type="tr:LanguageFeatureList"
               minOccurs="0"
               maxOccurs="unbounded" />
  </xs:sequence>
</xs:complexType>
```

| tr:Language Metadata Elements | | |
|---|---|---|
| Element | Definition | |
| name | *Value type:* | a prefixless XML name |
| | *Semantic Meaning:* | The name of the language without a version suffix. |
| | *Occurrences:* | required |
| version | *Value type:* | a string with optional attributes |

| tr:Language Metadata Elements | | |
|---|---|---|
| **Element** | **Definition** | |
| | Semantic Meaning: | A version of the language supported by the server. |
| | Occurrences: | required; multiple occurrences allowed. |
| description | Value type: | string: xs:token |
| | Semantic Meaning: | A short, human-readable description of the query language. |
| | Occurrences: | optional |
| languageFeatures | Value type: | composite: tr:LanguageFeatureList |
| | Semantic Meaning: | Optional features of the query language, grouped by feature type. |
| | Occurrences: | optional; multiple occurrences allowed. |
| | Comments: | This includes listing user defined functions, geometry support, or similar concepts. |

tr:Version Type Schema Definition

```
<xs:complexType name="Version" >
  <xs:simpleContent >
    <xs:extension base="xs:token" >
      <xs:attribute name="ivo-id" type="xs:anyURI" />
    </xs:extension>
  </xs:simpleContent>
</xs:complexType>
```

| tr:Version Attributes | | |
|---|---|---|
| **Attribute** | **Definition** | |
| ivo-id | Value type: | a URI: xs:anyURI |
| | Semantic Meaning: | An optional IVORN of the language. |
| | Occurrences: | optional |
| | Comments: | To more formally define a language supported by a service, a resource record for the language can be created, either centrally on the Registry of Registries or by other registry operators. When such a record exists, the language element's ivo-id should point to it. |

Query languages may support optional features. For ADQL, the most prominent of those are user-defined functions, i.e., functions not defined in the language standard but added by the operators of the service, and geometry functions. Such optional features may be communicated to the service client in `tr:languageFeatures` elements.

Each such list is labelled with a `type` attribute indicating the type of language option being described. This string should be an IVORN whose semantics in this context, along with the semantics of the content of its descendant `feature/form` elements, can be documented in association with the language in question.

TAPRegExt itself defines the following feature types:

`ivo://ivoa.net/std/TAPRegExt#features-udf`

Each feature declares a user-defined ADQL (or similar) function supported. The content of the `form` element must be the signature of the function, written to match the `signature` nonterminal in the following grammar:

```
signature ::= <funcname> <arglist> "->" <type_name>
funcname ::= <regular_identifier>
arglist ::= "(" <arg> { "," <arg> } ")"
arg ::= <regular_identifier> <type_name>
```

The `type_name` nonterminal is not defined by the ADQL grammar. For the purposes of TAPRegExt, it is sufficient to assume it expands to "some sort of SQL type specifier" (which may include spaces and parentheses). For an enumeration of common types in ADQL, refer to the last column of the table in section 2.5 of [TAP].

Example:

```
<languageFeatures type="ivo://ivoa.net/std/TAPRegExt#features-udf">
  <feature>
    <form>match(pattern TEXT, string TEXT) -> INTEGER</form>
    <description>
      match returns 1 if the POSIX regular expression pattern
      matches anything in string, 0 otherwise.
    </description>
  </feature>
</languageFeatures>
```

`ivo://ivoa.net/std/TAPRegExt#features-adqlgeo`

Each feature declares support for one of the geometry functions defined by ADQL (support for these functions is in general optional for ADQL implementations, though TAP imposes some constraints on what combinations of support are permitted).

The signature of these functions, where supported, is fixed by ADQL; the content of the `form` element is just the name of the function.

Example:

```
<feature>
  <form>CONTAINS</form>
</feature>
```

tr:LanguageFeatureList Type Schema Definition

```
<xs:complexType name="LanguageFeatureList" >
  <xs:sequence >
    <xs:element name="feature" type="tr:LanguageFeature" minOccurs="0"
              maxOccurs="unbounded" />
  </xs:sequence>
  <xs:attribute name="type" type="xs:anyURI" use="required" />
</xs:complexType>
```

| tr:LanguageFeatureList Metadata Elements | |
|---|---|
| Element | Definition |
| feature | *Value type:* composite: tr:LanguageFeature |
| | *Semantic Meaning:* A language feature of the type given by this element's type attribute. |
| | *Occurrences:* optional; multiple occurrences allowed. |

| tr:LanguageFeatureList Attributes | |
|---|---|
| Attribute | Definition |
| type | *Value type:* a URI: xs:anyURI |
| | *Semantic Meaning:* The type of the features given here. |
| | *Occurrences:* required |
| | *Comments:* This is in general an IVORN. TAPRegExt itself gives IVORNs for defining user defined functions and geometry support. |

tr:LanguageFeature Type Schema Definition

```
<xs:complexType name="LanguageFeature" >
  <xs:sequence >
    <xs:element name="form" type="xs:token" />
    <xs:element name="description" type="xs:string" minOccurs="0" />
  </xs:sequence>
</xs:complexType>
```

| tr:LanguageFeature Metadata Elements | |
|---|---|
| Element | Definition |
| form | *Value type:* string: xs:token |
| | *Semantic Meaning:* Formal notation for the language feature. |
| | *Occurrences:* required |
| | *Comments:* The syntax for the content of this element is defined by the type attribute of its parent language list. |
| description | *Value type:* string: xs:string |
| | *Semantic Meaning:* Human-readable freeform documentation for the language feature. |
| | *Occurrences:* optional |

## 2.4. Output Formats

A TAP service may offer a variety of output formats. What output formats are available is defined using `out-putFormat` elements. They declare a MIME type [RFC2045] as well as aliases (the shorthand forms the server also accepts in the FORMAT parameter). If desired, the format can be further described with an IVORN in the ivo-id attribute; TAPRegExt provides keys for some variants of VOTables which are not interoperably distinguishable by their MIME types so far:

```
output-votable-td
```
A VOTable in which all DATA elements contain a TABLEDATA element
```
output-votable-binary
```
A VOTable in which all DATA elements contain a STREAM element with a BINARY child
```
output-votable-binary2
```
A VOTable in which all DATA elements contain a STREAM element with a child of the yet-to-be-defined BINARY2 VOTable element

tr:OutputFormat Type Schema Definition

```
<xs:complexType name="OutputFormat" >
  <xs:sequence >
    <xs:element name="mime" type="xs:token" />
    <xs:element name="alias" type="xs:token" minOccurs="0"
                maxOccurs="unbounded" />
  </xs:sequence>
  <xs:attribute name="ivo-id" type="xs:anyURI" />
</xs:complexType>
```

| tr:OutputFormat Metadata Elements | | |
|---|---|---|
| Element | Definition | |
| mime | Value type: | string: xs:token |
| | Semantic Meaning: | The MIME type of this format. |
| | Occurrences: | required |
| | Comments: | The format of this string is specified by RFC 2045. The service has to accept this string as a value of the FORMAT parameter. |
| alias | Value type: | string: xs:token |
| | Semantic Meaning: | Other values of FORMAT ("shorthands") that make the service return documents with the MIME type. |
| | Occurrences: | optional; multiple occurrences allowed. |

| tr:OutputFormat Attributes | | |
|---|---|---|
| Attribute | Definition | |
| ivo-id | Value type: | a URI: xs:anyURI |
| | Semantic Meaning: | An optional IVORN of the output format. |
| | Occurrences: | optional |
| | Comments: | When the MIME type does not uniquely define the format (or a generic MIME like application/octet-stream or text/plain is given), the IVORN can point to a key or StandardsRegExt document defining the format more precisely. To see values defined in TAPRegExt, retrieve the ivo://ivoa.net/std/TAPRegExt resource record and look for keys starting with "output-". |

## 2.5. Upload Methods

TAP services should allow the upload of VOTables. They can support various methods to do this: HTTP upload, retrieval by URL, but also VOSpace or possibly retrieval using Grid protocols. Since an actual specification of the details of such protocols is far beyond the scope of a registry document and probably would not benefit clients anyway, the upload methods are given as IVORNs.

IVORNs for the standard upload methods are provided within the resource record `ivo://ivoa.net/std/TAPRegExt`. The IVORNs are built by using the keys as fragment identifiers within the TAPRegExt IVORN.

It is permitted to register upload methods under authorities other than ivoa.net. The registry records can then provide more in-depth information. For the upload methods defined in the TAP specification, however, the IVORNs of the keys in the TAPRegExt resource record must be used to enable clients to identify supported methods using string comparisons.

This document defines the following protocol identifiers:

- `upload-inline` -- HTTP upload as per section 2.5.2 of [TAP].
- `upload-http` -- retrieval from an http URL.
- `upload-https` -- retrieval from an https URL.
- `upload-ftp` -- retrieval from an ftp URL.

Thus, a service offering upload by retrieving from ftp and http URLs would say:

```
<uploadMethod ivo-id="ivo://ivoa.net/std/TAPRegExt#upload-http"/>
<uploadMethod ivo-id="ivo://ivoa.net/std/TAPRegExt#upload-ftp"/>
```

tr:UploadMethod Type Schema Definition

```
<xs:complexType name="UploadMethod" >
  <xs:complexContent >
    <xs:restriction base="xs:anyType" >
      <xs:attribute name="ivo-id" type="xs:anyURI" />
    </xs:restriction>
  </xs:complexContent>
</xs:complexType>
```

| tr:UploadMethod Attributes | | |
|---|---|---|
| Attribute | Definition | |
| ivo-id | *Value type:* | a URI: xs:anyURI |
| | *Semantic Meaning:* | The IVORN of the upload method. |
| | *Occurrences:* | optional |

## 2.6. Resource Limits

TAP services usually impose certain limits on resource usage by clients, e.g., a maximum run time per query, or a maximum number of rows in the result set. Services assign such limits to newly created jobs and may allow raising the limits by means of queries or query parameters (e.g., the size of the result set is limited by the `MAXREC` parameter, whereas the date of job destruction may be changed by posting to the `destruction` parameter). Services may put some limit to how far the resource limitations may be raised.

TAPRegExt's capabilities element allows the declaration of such limits. These declarations are primarily intended for human consumption and should give conservative guidelines. Thus, the operators of a service implementing a complex, possibly dynamic limits policy should give lower estimates here.

If a service supports authentication and has different limits depending on what user is authenticated, it should return the limits applying to the logged user.

The resource limits applying to newly created jobs are given in `default` elements, the limits beyond which users cannot raise the limits are given in `hard` elements.

Note that the absence of a specification of limits does not imply that no limits are enforced.

### Limits on Time

This document defines two time-like resource limits:

- `retentionPeriod` -- the time from job creation until `destruction`.
- `executionDuration` -- the maximal run time given to a query.

All values in time-like limits are given in seconds. Both `retentionPeriod` and `executionDuration` are of type `tr:TimeLimits`.

tr:TimeLimits Type Schema Definition

```
<xs:complexType name="TimeLimits" >
  <xs:sequence >
    <xs:element name="default" type="xs:integer" minOccurs="0" maxOccurs="1" />
    <xs:element name="hard" type="xs:integer" minOccurs="0" maxOccurs="1" />
  </xs:sequence>
</xs:complexType>
```

| tr:TimeLimits Metadata Elements | | |
|---|---|---|
| Element | Definition | |
| default | Value type: | integer |
| | Semantic Meaning: | The value of this limit for newly-created jobs, given in seconds. |
| | Occurrences: | optional |
| hard | Value type: | integer |
| | Semantic Meaning: | The value this limit cannot be raised above, given in seconds. |
| | Occurrences: | optional |

### Limits on Data

Limits on data are expressed much like time limits in that they give `default` and a `hard` value as well. Both those values have a unit attribute that can either be `byte` or `row` for data limits.

This document defines two resource limits on data:

- `outputLimit` -- if `unit` is `row` here, the `default` gives the value of TAP's `MAXREC` parameter the service will use when none is specified.

- `uploadLimit` -- the maximum size of uploads. This is not a TAP adjustable parameter. The `default` value advises clients about the server's wishes as to a limit above which some sort of acknowledgement should be requested from the user. The `hard` limit gives the maximum size of an upload to the server.

Data limits are defined using the `tr:DataLimits` and `tr:DataLimit` types:

tr:DataLimits Type Schema Definition

```
<xs:complexType name="DataLimits" >
  <xs:sequence >
    <xs:element name="default" type="tr:DataLimit" minOccurs="0" maxOccurs="1" />
    <xs:element name="hard" type="tr:DataLimit" minOccurs="0" maxOccurs="1" />
  </xs:sequence>
</xs:complexType>
```

| tr:DataLimits Metadata Elements | | |
|---|---|---|
| Element | Definition | |
| default | *Value type:* | an integer with optional attributes |
| | *Semantic Meaning:* | The value of this limit for newly-created jobs. |
| | *Occurrences:* | optional |
| hard | *Value type:* | an integer with optional attributes |
| | *Semantic Meaning:* | The value this limit cannot be raised above. |
| | *Occurrences:* | optional |

tr:DataLimit Type Schema Definition

```
<xs:complexType name="DataLimit" >
  <xs:simpleContent >
    <xs:extension base="xs:integer" >
      <xs:attribute name="unit" use="required" >
        <xs:simpleType >
          <xs:restriction base="xs:token" >
            <xs:enumeration value="byte" />
            <xs:enumeration value="row" />
          </xs:restriction>
        </xs:simpleType>
      </xs:attribute>
    </xs:extension>
  </xs:simpleContent>
</xs:complexType>
```

| tr:DataLimit Attributes | | |
|---|---|---|
| Attribute | Definition | |
| unit | *Value type:* | string with controlled vocabulary |
| | *Semantic Meaning:* | The unit of the limit specified. |
| | *Occurrences:* | required |

| tr:DataLimit Attributes | |
|---|---|
| Attribute | Definition |
| | *Allowed Values:* `byte`, `row` |

## 2.7. The Capability Record

Using the types defined above, the `tr:TableAccess` type can be defined. Note that it is a type, not a (global) element. In instance documents, you will typically place it in a capability element with an explicit type specification, like this:

```
<capability
  xmlns:tr="http://www.ivoa.net/xml/TAP/v1.0"
  xmlns:xsi="http://www.w3.org/2001/XMLSchema-instance"
  standardID="ivo://ivoa.net/std/TAP"
  xsi:type="tr:TableAccess">
  ...
```

By restriction from VOResource's `vr:Capability`, the `standardID` attribute of `tr:TableAccess`-typed capabilities is fixed to `ivo://ivoa.net/std/TAP` in this version. This string can be used to locate TAP services in the registry.

tr:TableAccess Type Schema Definition

```
<xs:complexType name="TableAccess" >
  <xs:complexContent >
    <xs:extension base="tr:TAPCapRestriction" >
      <xs:sequence >
        <xs:element name="dataModel" type="tr:DataModelType" minOccurs="0"
                    maxOccurs="unbounded" />
        <xs:element name="language" type="tr:Language" minOccurs="1"
                    maxOccurs="unbounded" />
        <xs:element name="outputFormat" type="tr:OutputFormat" minOccurs="1"
                    maxOccurs="unbounded" />
        <xs:element name="uploadMethod" type="tr:UploadMethod" minOccurs="0"
                    maxOccurs="unbounded" />
        <xs:element name="retentionPeriod" type="tr:TimeLimits" minOccurs="0"
                    maxOccurs="1" />
        <xs:element name="executionDuration" type="tr:TimeLimits" minOccurs="0"
                    maxOccurs="1" />
        <xs:element name="outputLimit" type="tr:DataLimits" minOccurs="0"
                    maxOccurs="1" />
        <xs:element name="uploadLimit" type="tr:DataLimits" minOccurs="0"
                    maxOccurs="1" />
      </xs:sequence>
    </xs:extension>
  </xs:complexContent>
</xs:complexType>
```

| tr:TableAccess Extension Metadata Elements | | |
|---|---|---|
| Element | Definition | |
| dataModel | *Value type:* | a string with optional attributes |
| | *Semantic Meaning:* | Identifier of IVOA-approved data model supported by the service. |
| | *Occurrences:* | optional; multiple occurrences allowed. |

| tr:TableAccess Extension Metadata Elements | |
|---|---|
| Element | Definition |
| language | **Value type:** composite: tr:Language<br>**Semantic Meaning:** Language supported by the service.<br>**Occurrences:** required; multiple occurrences allowed. |
| outputFormat | **Value type:** composite: tr:OutputFormat<br>**Semantic Meaning:** Output format supported by the service.<br>**Occurrences:** required; multiple occurrences allowed. |
| uploadMethod | **Value type:** composite: tr:UploadMethod<br>**Semantic Meaning:** Upload method supported by the service.<br>**Occurrences:** optional; multiple occurrences allowed.<br>**Comments:** The absence of upload methods indicates that the service does not support uploads at all. |
| retentionPeriod | **Value type:** composite: tr:TimeLimits<br>**Semantic Meaning:** Limits on the time between job creation and destruction time.<br>**Occurrences:** optional |
| executionDuration | **Value type:** composite: tr:TimeLimits<br>**Semantic Meaning:** Limits on executionDuration.<br>**Occurrences:** optional |
| outputLimit | **Value type:** composite: tr:DataLimits<br>**Semantic Meaning:** Limits on the size of data returned.<br>**Occurrences:** optional |
| uploadLimit | **Value type:** composite: tr:DataLimits<br>**Semantic Meaning:** Limits on the size of uploaded data.<br>**Occurrences:** optional |

## A. The Full Schema

```
<xs:schema version="1.0" targetNamespace="http://www.ivoa.net/xml/TAPRegExt/v1.0"
elementFormDefault="unqualified" attributeFormDefault="unqualified" xmlns:xml="http://
www.w3.org/XML/1998/namespace" xmlns:xs="http://www.w3.org/2001/XMLSchema" xmlns:vr="http://
www.ivoa.net/xml/VOResource/v1.0" xmlns:vm="http://www.ivoa.net/xml/VOMetadata/v0.1"
xmlns:tr="http://www.ivoa.net/xml/TAPRegExt/v1.0" xmlns:xsi="http://www.w3.org/2001/XM-
LSchema-instance">
  <xs:annotation>
    <xs:appinfo>
```

```xml
      <vm:schemaName>TAPRegExt</vm:schemaName>
      <vm:schemaPrefix>xs</vm:schemaPrefix>
      <vm:targetPrefix>tr</vm:targetPrefix>
    </xs:appinfo>
    <xs:documentation> A description of the capabilities metadata for TAP services.
    </xs:documentation>
  </xs:annotation>
  <xs:import namespace="http://www.ivoa.net/xml/VOResource/v1.0" schemaLocation="http://
  www.ivoa.net/xml/VOResource/VOResource-v1.0.xsd"/>
  <xs:complexType name="TAPCapRestriction" abstract="true">
    <xs:annotation>
      <xs:documentation> An abstract capability that fixes the standardID to the IVOA ID for
      the TAP standard. </xs:documentation>
      <xs:documentation> See vr:Capability for documentation on inherited children.
      </xs:documentation>
    </xs:annotation>
    <xs:complexContent>
      <xs:restriction base="vr:Capability">
        <xs:sequence>
          <xs:element name="validationLevel" type="vr:Validation" minOccurs="0"
          maxOccurs="unbounded"/>
          <xs:element name="description" type="xs:token" minOccurs="0"/>
          <xs:element name="interface" type="vr:Interface" minOccurs="0" maxOccurs="unbounded"/>
        </xs:sequence>
        <xs:attribute name="standardID" type="vr:IdentifierURI" use="required" fixed="ivo://
        ivoa.net/std/TAP"/>
      </xs:restriction>
    </xs:complexContent>
  </xs:complexType>
  <xs:complexType name="TableAccess">
    <xs:annotation>
      <xs:documentation> The capabilities of a TAP server. </xs:documentation>
      <xs:documentation> The capabilities attempt to define most issues that the TAP standard
      leaves to the implementors ("may", "should"). </xs:documentation>
    </xs:annotation>
    <xs:complexContent>
      <xs:extension base="tr:TAPCapRestriction">
        <xs:sequence>
          <xs:element name="dataModel" type="tr:DataModelType" minOccurs="0"
          maxOccurs="unbounded">
            <xs:annotation>
              <xs:documentation> Identifier of IVOA-approved data model supported by the service.
              </xs:documentation>
            </xs:annotation>
          </xs:element>
          <xs:element name="language" type="tr:Language" minOccurs="1" maxOccurs="unbounded">
            <xs:annotation>
              <xs:documentation> Language supported by the service. </xs:documentation>
            </xs:annotation>
          </xs:element>
          <xs:element name="outputFormat" type="tr:OutputFormat" minOccurs="1"
          maxOccurs="unbounded">
            <xs:annotation>
              <xs:documentation> Output format supported by the service. </xs:documentation>
            </xs:annotation>
          </xs:element>
          <xs:element name="uploadMethod" type="tr:UploadMethod" minOccurs="0"
          maxOccurs="unbounded">
            <xs:annotation>
              <xs:documentation> Upload method supported by the service. </xs:documentation>
              <xs:documentation> The absence of upload methods indicates that the service does
              not support uploads at all. </xs:documentation>
            </xs:annotation>
          </xs:element>
          <xs:element name="retentionPeriod" type="tr:TimeLimits" minOccurs="0" maxOccurs="1">
            <xs:annotation>
```

```xml
          <xs:documentation> Limits on the time between job creation and destruction time.
          </xs:documentation>
        </xs:annotation>
      </xs:element>
      <xs:element name="executionDuration" type="tr:TimeLimits" minOccurs="0" maxOccurs="1">
        <xs:annotation>
          <xs:documentation> Limits on executionDuration. </xs:documentation>
        </xs:annotation>
      </xs:element>
      <xs:element name="outputLimit" type="tr:DataLimits" minOccurs="0" maxOccurs="1">
        <xs:annotation>
          <xs:documentation> Limits on the size of data returned. </xs:documentation>
        </xs:annotation>
      </xs:element>
      <xs:element name="uploadLimit" type="tr:DataLimits" minOccurs="0" maxOccurs="1">
        <xs:annotation>
          <xs:documentation> Limits on the size of uploaded data. </xs:documentation>
        </xs:annotation>
      </xs:element>
    </xs:sequence>
  </xs:extension>
 </xs:complexContent>
</xs:complexType>
<xs:complexType name="DataModelType">
  <xs:annotation>
    <xs:documentation> An IVOA defined data model, identified by an IVORN intended for ma-
    chine consumption and a short label intended for human comsumption. </xs:documentation>
  </xs:annotation>
  <xs:simpleContent>
    <xs:extension base="xs:token">
      <xs:attribute name="ivo-id" type="vr:IdentifierURI" use="required">
        <xs:annotation>
          <xs:documentation> The IVORN of the data model. </xs:documentation>
        </xs:annotation>
      </xs:attribute>
    </xs:extension>
  </xs:simpleContent>
</xs:complexType>
<xs:complexType name="Language">
  <xs:annotation>
    <xs:documentation> A query language supported by the service. </xs:documentation>
    <xs:documentation> Each language element can describe one or more versions of a language.
    Either name alone or name-version can be used as values for the server's LANG parameter.
    </xs:documentation>
  </xs:annotation>
  <xs:sequence>
    <xs:element name="name" type="xs:NCName">
      <xs:annotation>
        <xs:documentation> The name of the language without a version suffix.
        </xs:documentation>
      </xs:annotation>
    </xs:element>
    <xs:element name="version" type="tr:Version" minOccurs="1" maxOccurs="unbounded">
      <xs:annotation>
        <xs:documentation> A version of the language supported by the server.
        </xs:documentation>
      </xs:annotation>
    </xs:element>
    <xs:element name="description" type="xs:token" minOccurs="0">
      <xs:annotation>
        <xs:documentation> A short, human-readable description of the query language.
        </xs:documentation>
      </xs:annotation>
    </xs:element>
    <xs:element name="languageFeatures" type="tr:LanguageFeatureList" minOccurs="0"
    maxOccurs="unbounded">
      <xs:annotation>
```

```xml
      <xs:documentation> Optional features of the query language, grouped by feature type.
      </xs:documentation>
      <xs:documentation> This includes listing user defined functions, geometry support, or
      similar concepts. </xs:documentation>
    </xs:annotation> </xs:documentation>
   </xs:element>
  </xs:sequence>
</xs:complexType>
<xs:complexType name="Version">
 <xs:annotation>
  <xs:documentation> One version of the language supported by the service.
  </xs:documentation>
  <xs:documentation> If the service supports more than one version of the language, include
  multiple version elements. It is recommended that you use a version numbering scheme like
  MAJOR.MINOR in such a way that sorting by ascending character codes will leave the most
  recent version at the bottom of the list. </xs:documentation>
 </xs:annotation>
 <xs:simpleContent>
  <xs:extension base="xs:token">
    <xs:attribute name="ivo-id" type="xs:anyURI">
     <xs:annotation>
       <xs:documentation> An optional IVORN of the language. </xs:documentation>
       <xs:documentation> To more formally define a language supported by a service, a re-
       source record for the language can be created, either centrally on the Registry of
       Registries or by other registry operators. When such a record exists, the language
       element's ivo-id should point to it. </xs:documentation>
     </xs:annotation>
    </xs:attribute>
  </xs:extension>
 </xs:simpleContent>
</xs:complexType>
<xs:complexType name="LanguageFeatureList">
 <xs:annotation>
  <xs:documentation> An enumeration of non-standard or non-mandatory features of a specific
  type implemented by the language. </xs:documentation>
  <xs:documentation> A feature type is a language-dependent concept like "user defined
  function", "geometry support", or possibly "units supported". A featureList gives all
  features of a given type applicable for the service. Multiple featureLists are possible.
  All feature in a given list are of the same type. This type is declared using the manda-
  tory type attribute, the value of which will typically be an IVORN. To see values defined
  in TAPRegExt, retrieve the ivo://ivoa.net/std/TAPRegExt resource record and look for keys
  starting with "features-". </xs:documentation>
 </xs:annotation>
 <xs:sequence>
  <xs:element name="feature" type="tr:LanguageFeature" minOccurs="0" maxOccurs="unbounded">
    <xs:annotation>
      <xs:documentation> A language feature of the type given by this element's type at-
      tribute. </xs:documentation>
    </xs:annotation>
  </xs:element>
 </xs:sequence>
 <xs:attribute name="type" type="xs:anyURI" use="required">
   <xs:annotation>
     <xs:documentation> The type of the features given here. </xs:documentation>
     <xs:documentation> This is in general an IVORN. TAPRegExt itself gives IVORNs for defin-
     ing user defined functions and geometry support. </xs:documentation>
   </xs:annotation>
 </xs:attribute>
</xs:complexType>
<xs:complexType name="LanguageFeature">
 <xs:annotation>
  <xs:documentation> A non-standard or non-mandatory feature implemented by the language..
  </xs:documentation>
 </xs:annotation>
 <xs:sequence>
  <xs:element name="form" type="xs:token">
    <xs:annotation>
```

```xml
          <xs:documentation> Formal notation for the language feature. </xs:documentation>
          <xs:documentation> The syntax for the content of this element is defined by the type
          attribute of its parent language list. </xs:documentation>
        </xs:annotation>
      </xs:element>
      <xs:element name="description" type="xs:string" minOccurs="0">
        <xs:annotation>
          <xs:documentation> Human-readable freeform documentation for the language feature.
          </xs:documentation>
        </xs:annotation>
      </xs:element>
    </xs:sequence>
  </xs:complexType>
  <xs:complexType name="OutputFormat">
    <xs:annotation>
      <xs:documentation> An output format supported by the service. </xs:documentation>
      <xs:documentation> All TAP services must support VOTable output, preserving the MIME
      type of the input. Other output formats are optional. The primary identifier for an out-
      put format is the MIME type. If you want to register an output format, you must use
      a MIME type (or make one up using the x- syntax), although the concrete MIME syntax
      is not enforced by the schema. For more detailed specification, an IVORN may be used.
      </xs:documentation>
    </xs:annotation>
    <xs:sequence>
      <xs:element name="mime" type="xs:token">
        <xs:annotation>
          <xs:documentation> The MIME type of this format. </xs:documentation>
          <xs:documentation> The format of this string is specified by RFC 2045. The service has
          to accept this string as a value of the FORMAT parameter. </xs:documentation>
        </xs:annotation>
      </xs:element>
      <xs:element name="alias" type="xs:token" minOccurs="0" maxOccurs="unbounded">
        <xs:annotation>
          <xs:documentation> Other values of FORMAT ("shorthands") that make the service return
          documents with the MIME type. </xs:documentation>
        </xs:annotation>
      </xs:element>
    </xs:sequence>
    <xs:attribute name="ivo-id" type="xs:anyURI">
      <xs:annotation>
        <xs:documentation> An optional IVORN of the output format. </xs:documentation>
        <xs:documentation> When the MIME type does not uniquely define the format (or a generic
        MIME like application/octet-stream or text/plain is given), the IVORN can point to a key
        or StandardsRegExt document defining the format more precisely. To see values defined in
        TAPRegExt, retrieve the ivo://ivoa.net/std/TAPRegExt resource record and look for keys
        starting with "output-". </xs:documentation>
      </xs:annotation>
    </xs:attribute>
  </xs:complexType>
  <xs:complexType name="UploadMethod">
    <xs:annotation>
      <xs:documentation> An upload method as defined by IVOA. </xs:documentation>
      <xs:documentation> Upload methods are always identified by an IVORN. Descriptions can be
      obtained by dereferencing this IVORN. To see values defined in TAPRegExt, retrieve the
      ivo://ivoa.net/std/TAPRegExt resource record and look for keys starting with "upload-".
      You can register custom upload methods, but you must use the standard IVORNs for the up-
      load methods defined in the TAP specification. </xs:documentation>
    </xs:annotation>
    <xs:complexContent>
      <xs:restriction base="xs:anyType">
        <xs:attribute name="ivo-id" type="xs:anyURI">
          <xs:annotation>
            <xs:documentation> The IVORN of the upload method. </xs:documentation>
          </xs:annotation>
        </xs:attribute>
      </xs:restriction>
    </xs:complexContent>
```

```xml
      </xs:complexType>
      <xs:complexType name="TimeLimits">
        <xs:annotation>
          <xs:documentation> Time-valued limits, all values given in seconds. </xs:documentation>
        </xs:annotation>
        <xs:sequence>
          <xs:element name="default" type="xs:integer" minOccurs="0" maxOccurs="1">
            <xs:annotation>
              <xs:documentation> The value of this limit for newly-created jobs, given in seconds.
              </xs:documentation>
            </xs:annotation>
          </xs:element>
          <xs:element name="hard" type="xs:integer" minOccurs="0" maxOccurs="1">
            <xs:annotation>
              <xs:documentation> The value this limit cannot be raised above, given in seconds.
              </xs:documentation>
            </xs:annotation>
          </xs:element>
        </xs:sequence>
      </xs:complexType>
      <xs:complexType name="DataLimits">
        <xs:annotation>
          <xs:documentation> Limits on data sizes, given in rows or bytes. </xs:documentation>
        </xs:annotation>
        <xs:sequence>
          <xs:element name="default" type="tr:DataLimit" minOccurs="0" maxOccurs="1">
            <xs:annotation>
              <xs:documentation> The value of this limit for newly-created jobs. </xs:documentation>
            </xs:annotation>
          </xs:element>
          <xs:element name="hard" type="tr:DataLimit" minOccurs="0" maxOccurs="1">
            <xs:annotation>
              <xs:documentation> The value this limit cannot be raised above. </xs:documentation>
            </xs:annotation>
          </xs:element>
        </xs:sequence>
      </xs:complexType>
      <xs:complexType name="DataLimit">
        <xs:annotation>
          <xs:documentation> A limit on some data size, either in rows or in bytes.
          </xs:documentation>
        </xs:annotation>
        <xs:simpleContent>
          <xs:extension base="xs:integer">
            <xs:attribute name="unit" use="required">
              <xs:annotation>
                <xs:documentation> The unit of the limit specified. </xs:documentation>
              </xs:annotation>
              <xs:simpleType>
                <xs:restriction base="xs:token">
                  <xs:enumeration value="byte"/>
                  <xs:enumeration value="row"/>
                </xs:restriction>
              </xs:simpleType>
            </xs:attribute>
          </xs:extension>
        </xs:simpleContent>
      </xs:complexType>
  </xs:schema>
```

# B. Example Document

As an example, here is an instance document as it might be part of a response from a VOSI capability endpoint or embedded within a VOResource record:

```xml
<capability standardID="ivo://ivoa.net/std/TAP" xsi:type="tr:TableAccess" xmlns:xml="http://
www.w3.org/XML/1998/namespace" xmlns:tr="http://www.ivoa.net/xml/TAPRegExt/v1.0"
xmlns:vr="http://www.ivoa.net/xml/VOResource/v1.0" xmlns:vs="http://www.ivoa.net/xml/VO-
DataService/v1.1" xmlns:xsi="http://www.w3.org/2001/XMLSchema-instance">
  <interface role="std" xsi:type="vs:ParamHTTP">
    <accessURL use="base">http://localhost:8080/__system__/tap/run/tap</accessURL>
  </interface>
  <dataModel ivo-id="ivo://ivoa.net/std/ObsCore-1.0">ObsCore 1.0</dataModel>
  <language>
    <name>ADQL</name>
    <version ivo-id="ivo://ivoa.net/std/ADQL#v2.0">2.0</version>
    <description>ADQL 2.0</description>
    <languageFeatures type="ivo://ivoa.net/std/TAPRegExt#features-udf">
      <feature>
        <form>ivo_nocasematch(value TEXT, pattern TEXT) -> INTEGER</form>
        <description>ivo_nocasematch returns 1 if pattern matches value, 0 otherwise. pattern
        is defined as for the SQL LIKE operator, but the match is performed case-insensitively.
        This function in effect provides a surrogate for the ILIKE SQL operator that is missing
        from ADQL.</description>
      </feature>
      <feature>
        <form>ivo_hashlist_has(hashlist TEXT, item TEXT) -> INTEGER</form>
        <description>The function takes two strings; the first is a list of words not containing
        the hash sign (#), concatenated by hash signs, the second is a word not containing the
        hash sign. It returns 1 if, compared case-insensitively, the second argument is in the
        list of words coded in the first argument. The behaviour in case the the second argument
        contains a hash sign is unspecified.</description>
      </feature>
      <feature>
        <form>gavo_match(pattern TEXT, string TEXT) -> INTEGER</form>
        <description>gavo_match returns 1 if the POSIX regular expression pattern matches any-
        thing in string, 0 otherwise.</description>
      </feature>
      <feature>
        <form>ivo_hasword(haystack TEXT, needle TEXT) -> INTEGER</form>
        <description>gavo_hasword returns 1 if needle shows up in haystack, 0 otherwise. This is
        for "google-like"-searches in text-like fields. In word, you can actually employ a fair-
        ly complex query language; see http://www.postgresql.org/docs/8.3/static/textsearch.html
        for details.</description>
      </feature>
    </languageFeatures>
    <languageFeatures type="ivo://ivoa.net/std/TAPRegExt#features-adqlgeo">
      <feature>
        <form>BOX</form>
      </feature>
      <feature>
        <form>POINT</form>
      </feature>
      <feature>
        <form>CIRCLE</form>
      </feature>
      <feature>
        <form>POLYGON</form>
      </feature>
      <feature>
        <form>REGION</form>
      </feature>
      <feature>
        <form>CENTROID</form>
      </feature>
      <feature>
        <form>COORD1</form>
      </feature>
      <feature>
        <form>COORD2</form>
      </feature>
      <feature>
```

```xml
      <form>DISTANCE</form>
    </feature>
    <feature>
      <form>CONTAINS</form>
    </feature>
    <feature>
      <form>INTERSECTS</form>
    </feature>
    <feature>
      <form>AREA</form>
    </feature>
  </languageFeatures>
</language>
<outputFormat ivo-id="ivo://ivoa.net/std/TAPRegExt#output-votable-binary">
  <mime>text/xml</mime>
</outputFormat>
<outputFormat ivo-id="ivo://ivoa.net/std/TAPRegEXT#output-votable-td">
  <mime>application/x-votable+xml;encoding=tabledata</mime>
  <alias>votable/td</alias>
</outputFormat>
<outputFormat>
  <mime>text/html</mime>
  <alias>html</alias>
</outputFormat>
<outputFormat>
  <mime>application/fits</mime>
  <alias>fits</alias>
</outputFormat>
<outputFormat>
  <mime>text/csv</mime>
</outputFormat>
<outputFormat>
  <mime>text/csv;header=present</mime>
  <alias>csv</alias>
</outputFormat>
<outputFormat>
  <mime>text/tab-separated-values</mime>
  <alias>tsv</alias>
</outputFormat>
<outputFormat ivo-id="ivo://ivoa.net/std/TAPRegExt#output-votable-binary">
  <mime>application/x-votable+xml</mime>
  <alias>votable</alias>
</outputFormat>
<outputFormat>
  <mime>text/plain</mime>
</outputFormat>
<uploadMethod ivo-id="ivo://ivoa.net/std/TAPRegExt#upload-https"/>
<uploadMethod ivo-id="ivo://ivoa.net/std/TAPRegExt#upload-ftp"/>
<uploadMethod ivo-id="ivo://ivoa.net/std/TAPRegExt#upload-inline"/>
<uploadMethod ivo-id="ivo://ivoa.net/std/TAPRegExt#upload-http"/>
<retentionPeriod>
  <default>172800</default>
</retentionPeriod>
<executionDuration>
  <default>3600</default>
</executionDuration>
<outputLimit>
  <default unit="row">2000</default>
  <hard unit="row">20000000</hard>
</outputLimit>
<uploadLimit>
  <hard unit="byte">20000000</hard>
</uploadLimit>
</capability>
```

Note that the encoding parameter in the MIME type given for the output format `ivo://ivoa.net/std/TAPRegEXT#output-votable-td` in the example is not endorsed by IVOA. Within the example, it represents a local convention.

# Changes from Previous Versions

## Changes from WD-20110127

- userDefinedFunction was generalized to feature within languageFeatures.
- The uploadmethods StandardKeyEnumeration was replaced by a resource record for TAPRegExt as a whole. This now includes keys of output formats and features as well; therefore, upload method names in their new IVORNs are prefixed with upload-
- Schema version was bumped to 1.0 (yes, we indulge in unversioned schema changes before this becomes REC).
- uploadLimit interpretation was changed: The default limit is now "advisory" and to be interpreted as such by clients, the hard limit is what is actually required by the server.
- There's now an optional ivo-id attribute on the version element within language.
- There's now an optional ivo-id attribute on output formats.

## Changes from WD-20110727

- The namespace in the schema is now `http://www.ivoa.net/xml/TAPRegExt/v1.0` consistent with what has already been stated in the text.
- The IVORN for ADQL is now `ivo://ivoa.net/std/ADQL#v2.0`; it is defined here to be in ADQL's record since we do not want to wait for the ADQL standard to be fixed, but ADQL versioning should really not be done here, so a TAPRegExt IVORN is out of the question.
- The IVORN of the TAPRegExt standard is now `ivo://ivoa.net/std/TAPRegExt` to conform with other standard IVORNs. Unfortunately, this changes all other IVORNs dependent on this.
- We now allow AnyURI on the ivo-id of language to allow fragment identifiers as, e.g., in ADQL.

## Changes from PR-20120812

- Fixed units in limits to "row" and "byte".